\documentclass[prl,aps,twocolumn,preprintnumbers,amsmath,amssymb]{revtex4}
\begin{document}
\preprint{DFPD-13/TH/15}
\title{A new class of \boldmath{$N=1$} no-scale supergravity models}
\author{Gianguido Dall'Agata and Fabio Zwirner}
\affiliation{
Dipartimento di Fisica ed Astronomia `G.~Galilei', Universit\`a di Padova 
\\ 
and INFN, Sezione di Padova, Via Marzolo 8, I-35131 Padova, Italy}
\date{September 2, 2013}
%
%
\begin{abstract}
We introduce a new $N=1$ no-scale supergravity model with F- and D-term breaking.
It contains a single chiral supermultiplet ${\cal T}$ and a single U(1) vector multiplet $U$, gauging a non-anomalous axionic shift symmetry. 
Both supersymmetry and the gauge symmetry are spontaneously broken, with the spin-3/2, spin-1 and spin-1/2 masses sliding along a classical flat direction, with a single real massless scalar in the spectrum.
The other degrees of freedom are absorbed by the massive gravitino and vector.
We extend our model, under very mild conditions, to general gauge groups and matter content. 
\end{abstract}
%
\maketitle
%
%
\section{Introduction}

The two outstanding unsolved hierarchy problems in the physics of the fundamental interactions are the smallness of the present vacuum energy density and  of the Fermi scale of weak interactions  with respect to the Planck scale of gravitational interactions.
A motivated and realistic theoretical context where both problems can be addressed, although not completely solved, is $N=1$, $D=4$ supergravity coupled to gauge and matter multiplets~\cite{books}. 
Being non-renormalizable and non-unique, $N=1$ supergravity must be interpreted as the low-energy effective theory of a more fundamental theory, possibly string theory, which may eventually dictate its field content, defining functions and counterterms.
But the presence of (super-)gravitational interactions is the source of a negative semi-definite contribution to the scalar potential, which makes it possible, at least in principle, to decouple the vacuum energy from the supersymmetry-breaking scale. 

In generic supergravity models, breaking supersymmetry on a background sufficiently flat to be realistic requires a huge fine-tuning, already at the classical level.
A remarkable exception is provided by the so-called {\em no-scale} models \cite{cfkn}, where at the classical level the potential is positive semi-definite, supersymmetry is broken with vanishing vacuum energy on a continuum of degenerate vacua, and the gravitino mass, setting the supersymmetry-breaking scale in Minkowski space, slides along a flat direction. 
This leaves the hope that, if some special class of no-scale models can be found for which, with an appropriate ultraviolet completion, quantum corrections are particularly benign \cite{fkpzstr}, such quantum corrections could generate the desired hierarchies \cite{ekn}.

In all the $N=1$ no-scale models considered so far, supersymmetry breaking is entirely due to the auxiliary fields of the chiral multiplets (pure F-term breaking), and there is at least one complex flat direction at the classical level. 
The simplest and best known example is the original model of \cite{cfkn}. 
In the present letter, we introduce a new class of no-scale models with mixed F- and D-term breaking, and a single real flat direction of the classical potential. 
We begin by recalling the basic formalism of $N=1$ supergravity with vector and chiral multiplets, and the features of the no-scale models considered so far. 
We then introduce the simplest example of our new class of no-scale models,  where supersymmetry and a U(1) axionic gauge symmetry are both spontaneously broken at the same scale, sliding along a real flat direction corresponding to the only massless particle in the spectrum. 
We also show how the relation between the constant superpotential and the gauge coupling constant, which is essential for the no-scale properties, can be obtained by a consistent truncation from a one-parameter $N=2$ no-scale model.   
We conclude by showing how our simple model can be generalized, under very mild conditions, to include arbitrary gauge groups and chiral multiplet content, and comment on possible future developments. 

\section{Basics of \boldmath{$N=1$}, \boldmath{$D=4$} supergravity}

An $N=1$, $D=4$ supergravity model with chiral multiplets $\phi^i \sim (z^i, \psi^i)$ and vector multiplets $U^a \sim (\lambda^a, A_\mu^a)$ is specified by three ingredients \cite{books}. 
The first is the real and gauge-invariant function
\begin{equation}
G = K + \log |W|^2 \, ,
\label{ggen}
\end{equation}
where $K$ is the real K\"ahler potential and $W$ the holomorphic superpotential. 
We are not interested here in the gauging of the R-symmetry, leading to constant Fayet--Iliopoulos terms, thus we can assume that both $K$ and $W$ are gauge invariant.  
The second is the holomorphic gauge kinetic function $f_{ab}$. Generalized Chern--Simons terms may also be needed, 
but they will not play any r\^ole in this paper. 
The third are the holomorphic Killing vectors $X_a = X_a^i (z) (\partial/\partial z^i)$, which generate the analytic isometries of the K\"ahler manifold for the scalar fields that are gauged by the vector fields. 
In the following, for simplicity, we will always take $G$, $f_{ab}$ and $X_a$ as functions of the complex scalars $z^i$ rather than the superfields $\phi^i$.

The gauge transformation laws and covariant derivatives for the scalars in the chiral multiplets read
\begin{equation}
 \delta z^i  =  X_a^i \, \epsilon^a \, , 
\qquad
D_\mu z^i = \partial_\mu z^i - A^a_\mu X_a^i \, ,
\end{equation}
where $\epsilon^a$ are real parameters. The scalar potential is made of three contributions, controlled by the auxiliary fields of the gravitational, chiral and vector multiplets: 
$$
V = V_G + V_F + V_D \, , 
\quad
V_G = - 3 \, e^G  \le 0 \, ,
$$
\begin{equation}
 \label{vgen}
V_F = e^G  G^i G_i \ge 0 \, , 
\quad
V_D =  \frac12 D_a D^a \ge 0 \, .
\end{equation}
In the above equations, $e^G$ is the field-dependent gravitino mass term $m_{3/2}^2$, $G_i = \partial G / \partial z^i$, scalar field indices are raised with the inverse K\"ahler metric $G^{i \overline{k}}$, gauge indices are raised with $[(Re f)^{-1}]^{ab}$, and
\begin{equation}
 \label{eq:solD} 
D_a = i \, G_i \, X_a^i = i \, K_i \, X_a^i \, .  
\end{equation}
For a linearly realized gauge symmetry, $i \, K_i \, X_a^i = - K_i \, (T_a)^i_{\; k} z^k$, whilst for an axionic U(1) symmetry $X_a^i = i \, q_a^i$, where $q_a^i$ is a real constant, and we obtain the so-called field-dependent Fayet--Iliopoulos terms.
Notice that D-terms are actually proportional to F-terms, $F_i = e^{G/2} \, G_i$, which implies the well-known fact that there cannot be pure D-breaking of supergravity in Minkowski space.

\section{No-scale models with pure F-breaking}

The simplest no-scale model with pure F-breaking \cite{cfkn} contains just a chiral multiplet  ${\cal T} \sim (T , \widetilde{T})$, with K\"ahler potential
\begin{equation}
 \label{kaold}
K = - 3 \,  \log \left( T + \overline{T} \right) \, ,
\end{equation}
and a $T$-independent superpotential
\begin{equation}
 \label{wmod}
W = W_0 \, .
\end{equation}
Since $G^TG_T = 3$, $V=V_G + V_F =0$ and supersymmetry is broken with vanishing vacuum energy for any constant value of the massless complex scalar $T=t + i \, \tau$ ($t>0$). 
The Goldstino $\widetilde{T}$ is absorbed by the gravitino, with $m_{3/2}^2 = |W_0|^2/(8 \, t^3)$, so that $t$ plays the role of a `dilaton', setting the scale of the only non-vanishing mass term.

The model can be easily generalized to include additional chiral multiplets $\phi^{\widehat{k}}$ and vector multiplets $U^a$, as long as the equations $\langle G_{\widehat{k}} \rangle = \langle D_a \rangle = 0$ admit solutions.

No-scale models can be also considered, where several fields $\phi^\alpha$ take part in the exact cancellation between $V_G$ and $V_F$, thanks to the identity $G^\alpha G_\alpha = 3$. 
In such a case we can split the chiral multiplets as $\phi^i = (\phi^\alpha , \, \phi^{\widehat{k}})$, and again the no-scale properties are preserved as long as the equations $\langle G_{\widehat{k}} \rangle = \langle D_a \rangle = 0$ admit solutions.

\section{A new model with F- and D-breaking}

Consider a model with a vector multiplet $U  \sim (\lambda, A_\mu)$, a  chiral multiplet ${\cal T} \sim (T , \widetilde{T})$ and K\"ahler potential
\begin{equation}
 \label{kamod}
K = - 2 \,  \log \left( T + \overline{T} \right) \, ,
\end{equation}
where, in contrast with the previous section, $\tau$ is now an `axion' that shifts under the U(1) isometry gauged by the vector multiplet. 
The corresponding holomorphic Killing vector is just an imaginary constant,
\begin{equation}
 \label{kimod}
X^T = i \, q \, ,
\qquad
(q \in {\mathbb R}) \, .
\end{equation}
The most general form of the superpotential invariant under the gauged U(1) is then the one of (\ref{wmod}).
Notice that the gauged U(1) is not an R-symmetry, in contrast with similar models previously considered in \cite{cfgkm,vzprl}. 
Notice also that both fermions $\widetilde{T}$ and $\lambda$ are neutral under gauge transformations, therefore there cannot be pure gauge U(1)$^3$ or mixed U(1) anomalies unless we add other multiplets containing fermions with U(1) charges. 
For the gauge kinetic function, we take a positive real constant
\begin{equation}
 \label{fmod}
f = \frac{1}{g^2} \, .
\end{equation}
The scalar potential of (\ref{vgen}) is then the sum of
\begin{equation}
 \label{vgfdmod}
V_G = -  \frac{3 \, |W_0|^2}{4 \, t^2} \, , 
\quad
V_F =  \frac{|W_0|^2}{2 \, t^2} \, , 
\quad
V_D = \frac{g^2 \, q^2}{2 \, t^2} \, .
\end{equation}
As required by gauge invariance, $V$ does not depend on $\tau$: the axion is absorbed by the massive U(1) vector boson via the Higgs effect \cite{gsfour}. 
However, $V_G$, $V_F$ and $V_D$ all depend non-trivially on $t$.
Constant $W$, constant $f$ and the $(-2)$ factor multiplying the logarithm in $K$ are essential in ensuring that all three terms in (\ref{vgfdmod}) have the same $t$-dependence.
In particular, if we choose 
\begin{equation}
 \label{choice}
|W_0| = \sqrt{2} \, g \, |q| \, , 
\end{equation}
there is an exact cancellation and $V=V_G+V_F+V_D = 0$. 

The gauge symmetry and supersymmetry are broken on Minkowski space at all vacua, with the would-be Goldstone boson and fermion given by $\tau$ and by a linear combination of $\widetilde{T}$ and $\lambda$, respectively. The only massless particle in the spectrum is the real scalar $t$, and the non-vanishing squared masses are
 \begin{equation}
\label{spectrum}
m_{3/2}^2  = \frac{ g^2 \, q^2}{2 \, t^2}  \, ,
\quad
m_{1}^2  = \frac{ g^2 \, q^2}{t^2}  \, ,
\quad
m_{1/2}^2 = \frac{ g^2 \, q^2}{2 \, t^2}  \, ,
\end{equation}
for the gravitino, the vector, and the spin-1/2 fermion orthogonal to the would-be Goldstino, respectively.

Superficially, we may think that the choice of (\ref{choice}) is a fine-tuning. 
However, to argue that this is not the case we recall that $N=1$ superpotentials are often originated from the gauge interactions of some compactified higher-dimensional supergravity or superstring theory. 
Also, it is well known that in extended supergravities all potential terms arise from gauge interactions, and it is not difficult to find examples of $N=1$ F-term potentials arising from $N>1$ D-term potentials. 
To fully convince the reader, we now build an explicit $N=2$ no-scale model, with a single gauge coupling constant $g$, which admits a consistent truncation to our new $N=1$ no-scale model and explains the relation (\ref{choice}). 
 
 \section{\boldmath{$N=2$} to \boldmath{$N=1$} truncation} 
\label{sec:_n_2_to_n_1_truncation}

Inspired by \cite{Ferrara:1995gu}, we construct a simple $N=2$ no-scale model that, in addition to the gravity multiplet $\{g_{\mu\nu}, \psi_{\mu \, A}, A_\mu^0\}$, contains one vector multiplet $\{A_\mu^1, \lambda^A\}$ and one hypermultiplet $\{q^u, \zeta^\alpha\}$, charged with respect to a U(1) $\times$ U(1) gauge group ($A=1,2$; $u=0,\ldots,3$; $\alpha = 1,2$).
The complex scalar of the vector multiplet describes the SU(1,1)/U(1) manifold with prepotential
\begin{equation}
	F(X) = -\frac{i}{2}\left[(X^0)^2 - (X^1)^2\right] \,.
\end{equation}
Introducing $z \equiv X^1/X^0$, the associated K\"ahler potential is $K_V = - \log i\,[X^\Lambda \overline{F}_{\Lambda} - \overline{X}^\Lambda F_{\Lambda}] = - \log \left[2\left(1 - |z|^2\right)\right]$, where $F_{\Lambda} \equiv \partial_{\Lambda} F$, $(\Lambda=0,1)$.
The scalars of the hypermultiplet are described by the  Quaternionic-K\"ahler manifold SO(4,1)/SO(4), with metric
\begin{equation}
\label{quatmet}
	ds^2 = h_{u v}(q) dq^u dq^v = \frac{1}{2 \, (b^0)^2} \left(db^0 db^0 + db^x db^x\right) \, ,
\end{equation}
($x=1,2,3$), and SU(2) connection $\omega^x_y = \delta^x_y/b^0$.
The two vector fields gauge two of the three translational isometries of (\ref{quatmet}).
This produces, in $N=1$ normalization, a scalar potential \cite{Andrianopoli:1996cm}
\begin{equation}
\label{scalpot}
	\frac{V}{g^2} = 4 L^\Lambda \bar{L}^\Sigma k_{\Lambda}^u k_{\Sigma}^v h_{uv} + \left(g^{z \bar z} f_z^\Lambda\, \overline{f}_{\bar z}^\Sigma - 3 \overline{L}^\Lambda L^\Sigma\right) P_{\Lambda}^x P_{\Sigma}^x \, ,
\end{equation}
where $k_{\Lambda}^u$ are the Killing vectors of the gauged isometries, $P_{\Lambda}^x$ are the corresponding prepotentials, $L^\Lambda \equiv e^{K_V/2}X^\Lambda$ and $f_z^\Lambda \equiv e^{K_V/2} D_z X^\Lambda$.
We find that the potential vanishes identically for the choice
\begin{equation}
	k_0^u = q \,\delta^{u2}, \qquad k_1^u = q \, \delta^{u3} \, .
\end{equation}
In fact, the prepotentials satisfy $P_{\Lambda}^x = \omega_u^x k_{\Lambda}^u$ and the three terms in (\ref{scalpot}) become proportional to each other, so that $V = 0$.

We now consistently truncate this model from $N=2$ to $N=1$ following the procedure outlined in \cite{Andrianopoli:2001gm}.
All consistency conditions are fulfilled by imposing
\begin{equation}
\label{truncation}
	\psi_{\mu 2} = \zeta^1 = \lambda^{2} = A_\mu^0 = b^1 = b^2 = z = \epsilon_2 =0 \, ,
\end{equation}
so that in the $N=1$ model only the gravitational multiplet, one vector multiplet and one chiral multiplet survive.
Since the graviphoton is projected out, only the gauging of the $b^3$ shift symmetry survives in the truncated model.
This is consistent with the fact that the Quaternionic-K\"ahler manifold is truncated to the SU(1,1)/U(1) manifold with K\"ahler potential (\ref{kamod}), after the identifications $t=b^0$ and $\tau = b^3$.
Also, by applying (\ref{truncation}), the scalar field $z$ of the $N=2$ vector multiplet is projected out and the gauge kinetic matrix reduces to the constant value ${\cal N}|_{z=0} = diag \, (-i, -i)$, so that we can identify the $N=1$ gauge kinetic function as $f = -i/g^2\, \overline{\cal N}_{11}$.
We conclude that the resulting $N=1$ model is precisely the one presented in the previous section, where the $G$ function (\ref{ggen}) and the D-term (\ref{eq:solD})  can be identified adapting the general relations given in \cite{Andrianopoli:2001gm} to our conventions:
\begin{eqnarray}
 e^{K/2}W &=& g \, \left[L^0 P_0^2\right]_{z=b_1=b_2=0} =  \frac{g \, q}{\sqrt2\,t} \, , \\[2mm] 
 D &=& \left[P_3^2\right]_{z=b_1=b_2=0}= \frac{q}{t} \, .
\end{eqnarray}
It is easy to see that these expressions reproduce the $N=1$ model described in the previous section, including condition (\ref{choice}). Notice that, in contrast with the models considered in \cite{cvz}, for this $N=2$ model we have no choice on the number of $N=1$ chiral and vector multiplets surviving the truncation.


\section{Generalizations and discussion}

Our simple new $N=1$ no-scale model can be easily generalized. 

We have checked that, keeping the same field content and suitably adapting the gauge transformation properties when the gauged U(1) becomes an R-symmetry, the more general class of K\"ahler potentials $K = - p \log (T + \overline T) + \alpha + \beta (T + \overline{T})$, superpotentials $W=W_0 \, e^{- \gamma T}$, gauge kinetic functions $f= \delta + \epsilon \, T$,  inspired by \cite{gsfour,vzprl,fklp}, cannot generate no-scale models inequivalent to the one already discussed above, corresponding to $p=2$, $\alpha=\beta=\gamma=\epsilon=0$, $\delta=1/g^2$.

The inclusion of additional gauge multiplets is straightforward, and we can promote the gauge group from the U(1) gauged by the single vector multiplet $U$ to U(1) $\times {\cal G}$, gauged by the vector multiplets $U^a= (U,U^{\widehat{a}})$. We can also enlarge the set of chiral multiplets $\phi^i=({\cal T}, \phi^{\widehat{k}})$, as long as the $ \phi^{\widehat{k}}$ do not transform under the original U(1). 

To preserve the crucial features of our simple new no-scale model with F- and D-breaking, but make it more realistic, we can proceed as in the case of the no-scale models with pure F-breaking. We can extend the K\"ahler potential to 
\begin{equation}
 \label{kaext}
K = - 2 \,  \log \left( T + \overline{T} \right) + \Delta K (T+\overline{T},\phi^{\widehat{k}},\overline{\phi}^{\overline{\widehat{k}}}) \, ,
\end{equation}
the superpotential to
\begin{equation}
 \label{wext}
W = W_0 + \Delta W (\phi^{\widehat{k}}) \, ,
\end{equation}
and we can introduce a gauge kinetic function for the gauge group factor ${\cal G}$ of the form
\begin{equation}
 \label{fext}
 f_{\widehat{a} \widehat{b}}  = 
 f^{(0)} _{\widehat{a} \widehat{b}}   (\phi^{\widehat{k}}) 
 + 
 f^{(1)} _{\widehat{a} \widehat{b}}   (\phi^{\widehat{k}})  \ T \, .
\end{equation}
The conditions to be satisfied by the modifications (\ref{kaext})--(\ref{fext}) are that gauge invariance is preserved, at the classical and at the quantum level, that the equations  $\langle G_{\widehat{k}} \rangle = \langle D_{\widehat{a}} \rangle = 0$ admit solutions in field configuration space, and that $\langle \Delta W \rangle = \langle \Delta K \rangle = 0$ upon minimization.

For example, if we think of the $\phi^{\widehat{k}}$ and of the $U^{\widehat{a}}$ as the chiral and vector multiplets of some supersymmetric extension of the Standard Model, and we work in the approximation of small field fluctuations around $\langle  z^{\widehat{k}} \rangle = 0$, we can choose
\begin{equation}
 \label{karealext}
\Delta K  = \sum_{\widehat{k}} |  z^{\widehat{k}} |^2 (T+\overline{T})^{n_{\widehat{k}}} \, , 
\quad
(n_{\widehat{k}} \in \mathbb{Z}) \, ,
\end{equation}
\begin{equation}
 \label{wfrealext}
\Delta W  = \sum_{\widehat{k} \widehat{l} \widehat{m}} d_{\widehat{k} \widehat{l} \widehat{m}}
 z^{\widehat{k}}  z^{\widehat{l}}  z^{\widehat{m}} \, ,
 \qquad
f_{\widehat{a} \widehat{b}}  = 
\delta_{\widehat{a} \widehat{b}}  \ T  \, .
\end{equation}
Also in this case, for suitable values of the new parameters, there is a real valley of degenerate minima of the full potential $V$ that satisfy the consistency conditions mentioned above and keep the good features of the simple model. 
For the given choice, the supersymmetry-breaking scalar and gaugino masses for the additional chiral and vector supermultiplets would be:
\begin{equation}
 \label{hatmasses}
 \widetilde{m}_{\widehat{k}}^2 =  n_{\widehat{k}}  \,  m_{3/2}^2 \, , 
 \qquad
 M_{\widehat{a}}^2 = 4 \, m_{3/2}^2 \, . 
\end{equation}

Notice that in both simple no-scale models considered in this letter, the old and the new one, ${\rm Str} {\cal M}^2$ evaluated along the flat direction  is proportional to $m_{3/2}^2$ via an integer numerical constant.
Indeed, the spectrum in (\ref{spectrum}) gives ${\rm Str} {\cal M}^2 =0$, but for the reasons explained below we do not attach special importance to this. Having  ${\rm Str} {\cal M}^2 = n \, m_{3/2}^2$ ($n \in \mathbb{Z}$) leaves open the possibility that, once additional sectors are introduced to make the model realistic, with supersymmetry-breaking squared mass splittings also proportional to $m_{3/2}^2$, the condition ${\rm Str} {\cal M}^2 = 0$ can be fulfilled. 
This would ensure the absence of quadratically divergent one-loop corrections to the effective potential, which are the most serious sources of vacuum instability.  

The results of the present letter call for further investigations in more realistic models. For example, the results of the first run of the LHC suggest a little hierarchy between the mass scales of the Higgs, W and Z bosons on one side, the supersymmetric particles and the additional Higgs bosons on the other side. 
Can such a little hierarchy be embedded in a no-scale model with only two classical real flat directions, controlling the two mass scales above? 
Can we then generate both scales by dimensional transmutation, taking into account logarithmic quantum corrections in the effective supergravity and suitably parameterizing our ignorance of the corrections coming from its ultraviolet completion? 
We believe that the new no-scale models introduced in this paper provide new possibilities to address these intriguing questions, and work along these lines is in progress \cite{luo}. 
Finally, it would be interesting to understand whether our models can actually originate from string/M-theory compactifications with fluxes, since this would embed them in a context where all perturbative quantum corrections could be eventually calculable.
 
However, the above investigations go beyond the aim of the present letter and are left for future work.

%
\section{Acknowledgments}

We thank Sergio Ferrara, Hui Luo, Massimo `Nicolaus' Porrati and Giovanni Villadoro for discussions. This work was supported in part  by the ERC Advanced Grants no.226455 (\textit{SUPERFIELDS}) and no.267985 (\textit{DaMeSyFla}), by the European Programme PITN-GA-2009-237920  (\textit{UNILHC}), by the Padova University Project CPDA105015/10, by the MIUR grants RBFR10QS5J, PRIN 2009-KHZKRX and PRIN 2010YJ2NYW.
%
%

%
\end{document}